\documentclass[aps,prd,preprintnumbers,showpacs,showkeys,nofootinbib,
superscriptaddress,fleqn,floatfix,tightenlines,10pt]{revtex4-1}
\usepackage{amsmath,amsfonts,amssymb,amscd,amsxtra,amsthm}
\usepackage{graphicx}  
\usepackage{epstopdf}
\usepackage{dcolumn}  
\usepackage{bm}          
\usepackage{slashed}
\usepackage{cancel}
\usepackage{float}
\usepackage{mathtools}
\usepackage{amsbsy}
\usepackage{amstext}

\usepackage[utf8]{inputenc} 
\usepackage{booktabs}
\usepackage[normalem]{ulem} 
\usepackage[dvipsnames]{xcolor} 
\usepackage{tabularx}
\usepackage{enumitem}  
\usepackage{array} 
\usepackage{slashed}
\usepackage{tikz}
\usepackage{float}
\usepackage{multirow} 
\usepackage{cancel}
\renewcommand\sout{\bgroup \color{red} \ULdepth=-.5ex \ULset}

\makeatletter

\begin{document}  
\preprint{INHA-NTG-04/2021}
\title{Energy-momentum tensor of the nucleon on the light front: Abel
  tomography case} 
\author{June-Young Kim}
\email[E-mail: ]{Jun-Young.Kim@ruhr-uni-bochum.de}
\affiliation{Institut f\"ur Theoretische Physik II, Ruhr-Universit\"at
  Bochum, D-44780 Bochum, Germany}

\author{Hyun-Chul Kim}
\email[E-mail: ]{hchkim@inha.ac.kr}
\affiliation{Department of Physics, Inha University, Incheon 22212,
Republic of Korea}
\affiliation{School of Physics, Korea Institute for Advanced Study
  (KIAS), Seoul 02455, Republic of Korea}

\date{\today}
\begin{abstract}
We investigate the two-dimensional energy-momentum-tensor (EMT)
distributions of the nucleon on the light front, using the Abel
transforms of the three-dimensional EMT ones. We explicitly show that
the main features of all EMT distributions are kept intact in the
course of the Abel transform. We also examine the equivalence between
the global and local conditions for the nucleon stability in the
three-dimensional Breit frame and in the two-dimensional transverse
plane on the light front.  
We also discuss the two-dimensional force fields inside a nucleon on
the light front.  
\end{abstract}
\pacs{}
\keywords{Energy-momentum tensor, Nucleon, Abel transform,
  chiral-quark soliton model}
\maketitle
\section{Introduction\label{sec:1}}
A modern understanding of a hadronic form factor is based on the
generalized parton distributions (GPDs)~\cite{Mueller:1998fv, Ji:1996ek,
  Radyushkin:1996nd, Goeke:2001tz, Diehl:2003ny, Belitsky:2005qn}: For
example, the electromagnetic form factors of the nucleon can be viewed
as the first moments of the unpolarized vector GPD with respect to the
parton momentum fractions. Its second moments
give the nucleon gravitational form factors 
(GFFs)~\cite{Kobzarev:1962wt, Pagels:1966zza}, which are also defined
in terms of the nucleon matrix element of the energy-moment tensor
(EMT). Since the EMT arises from the response of the nucleon to a
change of the external space-time metric, the corresponding form
factors were christened as the GFFs. These GFFs furnish essential
information on the internal structure of the nucleon such as the mass
and spin distributions. The most nontrivial part of the GFFs comes
from the $D$-term (Druck-term) form factor~\cite{Polyakov:2002yz}, 
which is deeply related to the stability of the nucleon. It
reveals how the nucleon acquires its mechanical stability, which is
exhibited by the pressure and shear-force 
densities. These three-dimensional (3D) distributions are often
presented in the Breit frame (BF)~\cite{Polyakov:2002yz, Goeke:2007fp, 
  Polyakov:2018zvc}. In the meanwhile, there have been criticisms on
the validity of the 3D densities for the nucleon~\cite{Yennie:1957,
  Burkardt:2000za, Burkardt:2002hr, Belitsky:2005qn, Miller:2007uy,
  Miller:2010nz, Jaffe:2020ebz} since the experiments on the proton
structure performed by Hofstadter~\cite{Hofstadter:1956qs,
  Hofstadter:1956qs, Hofstadter:1957wk}. As pointed out in the
literatures, the 3D densities are only valid for nonrelativistic
particles such as atoms and nuclei, of which the intrinsic sizes are
much larger than the corresponding Compton wavelengths
($\lambdabar=\hbar/mc$). When it comes to the nucleon, however, the
ratio of the Compton wavelength to the radius is numerically of order
$\sim 1/4$, which may result in relativistic corrections up to 20~\%
for the distributions and up to 10~\% for the nucleon radii. Moreover,
the ratio is of order $1/N_c$, i.e. the relativistic corrections are 
parametrically small in the large $N_c$ limit of QCD, which we
employ in the present work.

The BF distributions for the static EMT can be interpreted
as quasiprobabilistic densities from the phase-space
perspective\cite{Lorce:2017wkb, Lorce:2018zpf, Lorce:2018egm,
  Lorce:2020onh, Lorce:2021gxs}. This  
means that the BF distributions, which are defined through the Wigner
distributions in the quantum phase space~\cite{Wigner:1932eb,
  Hillery:1983ms}, can be comprehended as quasiprobabilistic ones due
to Heisenberg's uncertainty relations.  On the other hand, if one
takes the infinite momentum frame (IMF), which indicates that the
nucleon is on the light-front (LF), the relativistic corrections 
to the distributions are suppressed kinematically and a transversely
localized state for the nucleon can be defined, which enables one to
construct the two-dimensional (2D) transverse distributions on the LF.  
They  provide a strict probabilistic
interpretation~\cite{Burkardt:2002hr, Miller:2007uy, Carlson:2007xd} 
and are not Lorentz contracted, since the initial and final states of
the nucleon lie on the mass shell.  Recently, it was shown in
Ref.~\cite{Lorce:2020onh} that the BF charge distributions can be
interpolated to the LF ones. In Ref.~\cite{Panteleeva:2021iip}, this
interpolation was applied to the EMT force distributions. This was
realized by the Abel transform~\cite{Abel} that had been used in the  
computerized medical tomography in connection with the Radon
transform~\cite{Natterer:2001}. Actually, the Abel transform was
already utilized in the deeply virtual Compton
scattering~\cite{Polyakov:2007rv, Moiseeva:2008qd}. As pointed out in
Ref.~\cite{Panteleeva:2021iip}, thus, it is of great importance to 
scrutinize how the EMT force distributions in the BF are related to
those in the IMF. The GFFs of the nucleon were already investigated in
the chiral quark-soliton model ($\chi$QSM)~\cite{Goeke:2007fp}. It was
shown how the nucleon acquires the stability by examining its pressure 
densities. The work was extended to the GFFs of the singly heavy
baryon $\Sigma_c$ in comparison with those of the 
nucleon~\cite{Kim:2020nug}, the force fields inside the nucleon and
$\Sigma_c$ being emphasized. Both the global and local stability
conditions~\cite{Perevalova:2016dln,Polyakov:2018zvc} for them were
carefully examined. In the present work, we want to study how the
mechanical densities obtained in Ref.~\cite{Kim:2020nug} can be
projected onto the 2D transverse plane by the Abel
transform. Consequently, we show explicitly that the global and local
stability conditions are equivalently satisfied both in the BF and on
the LF.   

The present work is organized as follows: In Sec.~\ref{sec:2}, we
define the GFFs and show how the mechanical densities are derived.
Then we relate the 3D distributions to the 2D ones by the Abel
transform. The 2D local stability conditions are presented. In
Sect.~\ref{sec:3}, we discuss the numerical results for both the 3D  
and 2D mechanical distributions such as the energy, spin, and pressure
and shear-force densities. We explicitly show that the global and
local stability conditions both in the BF and on the LF are equivalent
to each other. We finally visualize how the force fields are
distributed inside a nucleon in the transverse plane on the LF. 
The last Section summarizes the present results and
draws conclusions.    

\section{Gravitational form factors of the nucleon}
\label{sec:2}
The nucleon form factors of the symmetric EMT is
defined as~\cite{Kobzarev:1962wt,Pagels:1966zza}
\begin{align}
\langle p',\lambda' | \hat{T}^{\mu\nu}(0) | p, \lambda \rangle =
  \overline{u}_{\lambda'}(p') \left[ A(t) \frac{P^{\mu}P^{\nu}}{m} +
  J(t) \frac{i P^{\{\mu}\sigma^{\nu\}\alpha} \Delta_{\alpha}}{m} +
  \frac{D(t)}{4m} (\Delta^{\mu}\Delta^{\nu}-g^{\mu\nu} \Delta^{2})
  \right] u_{\lambda}(p)\, ,
\end{align}
where $\hat{T}^{\mu\nu}(0)$ denotes the symmetric EMT
 operator. The parenthesis
 $a^{\{\mu}b^{\nu\}}=(a^{\mu}b^{\nu}+a^{\nu}b^{\mu})/2$ stand for the   
 symmetrization operator. ${u}_{\lambda}(p)$ is the Dirac
spinor. Here, $\lambda (\lambda')$ 
represents the initial (final) spin projections. The normalization of
Dirac spinors is taken to be $\overline{u}_{\lambda'}(p)u_{\lambda}(p) = 2m
\delta_{\lambda'\lambda}$. We use the covariant normalization $\langle
p', \lambda'| p, \lambda \rangle = 2p^{0} (2\pi)^{3}\delta_{\lambda'
  \lambda} \delta^{(3)}(\bm{p}'-\bm{p})$ of one-particle states, and
introduce kinematical variables $P^{\mu}=(p^{\mu}+p'^{\mu})/2$,
$\Delta^{\mu}=p'^{\mu}-p^{\mu}$ and $\Delta^{2} = t$. The nucleon
matrix element of the EMT current is parametrized in terms of the
three real form factors, i.e., $A(t)$, $J(t)$, and
$D(t)$, which yield information on the mass, spin, and the stability
of the nucleon, respectively. The mass form factor $M(t)$ is obtained
in terms of these three form factors
\begin{align}
M(t) = A(t) - \frac{t}{4m^2} \bigg{(}A(t)
  -2J(t) +D(t)\bigg{)}.  
\label{eq:mass}
\end{align}
The mass $\varepsilon(r)$, angular momentum\footnote{The angular
  momentum distribution is decomposed in 
  terms of the monopole and quadrupole
  contributions~\cite{Polyakov:2002yz, Lorce:2018egm}, and both the
  distributions are not independent and interrelated
  as~\cite{Schweitzer:2019kkd}, i.e., $\rho^{\mathrm{quad}}_{J}(r)=
  -\frac{3}{2}\rho_{J}(r)$. The given distribution $\rho_{J}(r)$
  corresponds to the monopole contribution.} $\rho_{J}(r)$, pressure
$p(r)$ and shear force $s(r)$ distributions in the BF are derived by
the 3D inverse Fourier transform of the GFFs, i.e., $F=A, M, J, D$, in
terms of the multipole expansion~\cite{Lorce:2018egm, 
  Polyakov:2018zvc, Polyakov:2018rew, Kim:2020lrs}:
\begin{align}
\varepsilon(r) =  m \tilde{M}(r), \ \ 
\rho_{J}(r)= -\frac{1}{3}r \frac{d}{dr} \tilde{J}(r), \ \  s(r) =
  -\frac{1}{4m} r \frac{d}{dr} \frac{1}{r} \frac{d}{dr} \tilde{D}(r),
  \ \ p(r)=
  \frac{1}{6m}\frac{1}{r^{2}}\frac{d}{dr}r^{2}\frac{d}{dr}\tilde{D}(r), 
\label{eq:3DFF}
\end{align}
with the generic 3D Fourier transform 
\begin{align}
\tilde{F}(r) = \int \frac{d^{3}
  \bm{\Delta}}{(2\pi)^{3}}e^{-i\bm{\Delta} \cdot \bm{r}}
  F(-\bm{\Delta}^{2}) \ \  . 
\end{align}
As pointed out in Ref.~\cite{Yennie:1957}, The physical meaning of the
3D distributions is hampered by ambiguous relativistic corrections to
them. This ambiguity can be removed by considering the 
2D transverse distributions in the IMF~\cite{, 
  Burkardt:2000za, Burkardt:2002hr, Belitsky:2005qn, Miller:2007uy,
  Miller:2010nz, Jaffe:2020ebz}. 

Recently, the EMT distributions in
the IMF have been extensively explored in Refs.~\cite{Lorce:2018egm,
  Freese:2021czn, Lorce:2017wkb, Schweitzer:2019kkd, Lorce:2018egm,
  Panteleeva:2021iip, Freese:2021czn}. 
The corresponding 2D distributions for the momentum
$\varepsilon^{(2D)}(x_{\perp})$, angular momentum
$\rho_{J}^{(2D)}(x_{\perp})$, pressure $p^{(2D)}(x_{\perp})$ and shear
force $s^{(2D)}(x_{\perp})$ are obtained by the 2D inverse Fourier
transform 
\begin{align}
& \varepsilon^{(2D)}(x_{\perp}) =  P^{+} \tilde{A}(x_{\perp}), \ \ \ 
\rho^{(2D)}_{J}(x_{\perp})= -\frac{1}{2}x_{\perp} \frac{d}{dx_{\perp}}
  \tilde{J}(x_{\perp}),  \cr 
 & s^{(2D)}(x_{\perp}) = -\frac{1}{4P^{+}}
   x_{\perp}\frac{1}{dx_{\perp}} \frac{1}{x_{\perp}}
   \frac{d}{dx_{\perp}}  \tilde{D}(x_{\perp}), \ \ \
   p^{(2D)}(x_{\perp})=
   \frac{1}{8P^{+}}\frac{1}{x_{\perp}}
   \frac{d}{dx_{\perp}}x_{\perp}\frac{d}{dx_{\perp}}\tilde{D}(x_{\perp}),   
 \label{eq:2DFF}
\end{align}
where we define generically the 2D Fourier transform of the
corresponding GFFs as follows 
\begin{align}
\tilde{F}(x_{\perp}) = \int \frac{d^{2}
  \bm{\Delta}_{\perp}}{(2\pi)^{2}}e^{-i\bm{\Delta}_{\perp} \cdot
  \bm{x_{\perp}}} F(-\bm{\Delta}^{2}_{\perp}). 
\end{align}
The $\bm{x}_{\perp}$ and $\bm{\Delta}_{\perp}$ denote the
position and momentum vectors in the 2D plane transverse to the moving
direction of the nucleon, respectively. $P^{+}$ is the light-cone
momentum. Note that the mass, shear force, and pressure 
distributions are redefined by multiplying the Lorentz factors for
convenience~\cite{Panteleeva:2021iip} respectively as follows:  
\begin{align}
\mathcal{E}(x_\perp)= \frac{m}{P^{+}}\varepsilon^{(2D)}(x_\perp), \ \
  \mathcal{S}(x_\perp)= \frac{P^{+}}{2m}s^{(2D)}(x_\perp),  \ \
  \mathcal{P}(x_\perp)=\frac{P^{+}}{2m}p^{(2D)}(x_\perp) .
\label{eq:2DFFcov}
\end{align}
Since the longitudinal boost does not mix the longitudinal component
of the angular momentum, its distribution does not need to have an
additional Lorentz factor~\cite{Lorce:2017wkb}. The distributions in
the BF and on the LF, respectively given in Eq.~\eqref{eq:3DFF} and
Eq.~\eqref{eq:2DFFcov}, can be related to each other. Interestingly, the
connection between those distributions turns out to be the well-known 
Abel transforms as follows: 
\begin{align}
\left(1 -
  \frac{\partial^{2}_{(2D)}}{4m^{2}}\right)\mathcal{E}(x_{\perp}) &=
 2  \int^{\infty}_{x_{\perp}}
   \frac{rdr}{\sqrt{r^{2}-x^{2}_{\perp}}}
   \left[
   \varepsilon(r)
   +
   \frac{3}{2}p(r)
   +
   \frac{3}{2m}\frac{1}{r^{2}}\frac{d}{dr}r
   \rho_{J}(r)\right],
   \cr 
  \rho^{(2D)}_{J}(x_{\perp})&= 3 \int^{\infty}_{x_{\perp}}
  \frac{\rho_{J}(r)}{r}
  \frac{x^{2}_{\perp}dr}{\sqrt{r^{2}-x^{2}_{\perp}}}, \cr
\mathcal{S}(x_{\perp}) &=\int^{\infty}_{x_{\perp}} \frac{s(r)}{r}
  \frac{ x^{2}_{\perp}dr}{\sqrt{r^{2}-x^{2}_{\perp}}}, \cr
  \frac{1}{2}\mathcal{S}(x_{\perp})+\mathcal{P}(x_{\perp})
  &=\frac{1}{2}\int^{\infty}_{x_{\perp}}
  \left(\frac{2}{3}s(r)+p(r)\right) \frac{ r
  dr}{\sqrt{r^{2}-x^{2}_{\perp}}}. 
\label{eq:Abel_PS}
\end{align}
The Abel transform takes the same physical meaning from the 3D
distributions to the 2D ones. This means that the 2D distributions 
$\mathcal{E}(x_{\perp})$ and $\rho_{J}^{(2D)}(x_{\perp})$ take over
the meaning of the corresponding 3D ones as to how the mass and the
angular momentum of the baryon are distributed in the transverse
plane, respectively.   
Having integrated them over $\bm{x}_{\perp}$, we arrive at  
\begin{align}
\int d^{2}x_{\perp}\mathcal{E}(x_{\perp})=mA(0), \ \ \ \int
  d^{2}x_{\perp}\rho_{J}^{(2D)}(x_{\perp})=J(0),  
\label{eq:normal}
\end{align}
with the normalized form factor $A(0)=1$ and
$J(0)=1/2$, respectively. Being similar to the radii for the 3D
distributions, we can define respectively the 2D radii for the mass
and the angular momentum, which are relevant to the 3D radii as
\begin{align}
&\langle  x^{2}_{\perp} \rangle_{\mathrm{mass}} =\frac{1}{m}\int
  d^{2}x_{\perp} x^{2}_{\perp} \mathcal{E}(x_{\perp})
  =\frac{2}{3}\langle r^{2} \rangle_{\mathrm{mass}}
  +\frac{D(0)}{m^{2}}, \cr 
& \langle x^{2}_{\perp}
\rangle_{J}=2\int d^{2}x_{\perp} x^{2}_{\perp}
  \rho^{(2D)}_{J}(x_{\perp})= \frac{4}{5} \langle r^{2} \rangle_{J}.
\label{eq:radius}  
\end{align} 
The prefactors between the moments of the 2D and 3D distributions are
determined by the geometric factor given in 
Ref.~\cite{Panteleeva:2021iip} 
\begin{align}
\Omega_d=\frac{\sqrt{\pi}}{2}\frac{\Gamma\left(\frac{d+1}{2}\right)}{
  \Gamma\left(\frac{d+2}{2}\right)},
\end{align}
with dimension $d$.  

The conservation of the EMT current also provides the 2D stability 
condition of the nucleon. As in the 3D case, the 2D pressure and shear
force distributions are responsible for the stability condition. So,    
we obtain the  2D equilibrium equation from the conservation
of the EMT current, which is equivalent to the 3D expression as
follows: 
\begin{align}
p'(r)+ \frac{2s(r)}{r} +
  \frac{2}{3}s'(r)=0 \iff \mathcal{P}'(x_{\perp})+
  \frac{\mathcal{S}(x_{\perp})}{x_{\perp}} + 
  \frac{1}{2}\mathcal{S}'(x_{\perp})=0. 
\label{eq:diff_eq}
\end{align}
As shown in Eq.~\eqref{eq:diff_eq}, we find that the shear force and
pressure distributions are related to each other. Moreover, the 3D and
2D distributions should also comply respectively with the 3D and 2D
von Laue condition and those of its lower dimension subsystem in the
nucleon as    
\begin{align}
\int d^{3}r \, p(r)&=0 \iff
\int d^{2}{x}_{\perp} \mathcal{P}(x_{\perp})=0, \cr
 \int^{\infty}_{0} dr \, r  \left[p(r)-\frac{1}{3}s(r)\right]&=0 \iff
 \int^{\infty}_{0} d{x}_{\perp} \left[\mathcal{P}(x_{\perp})- 
  \frac{1}{2}\mathcal{S}(x_{\perp})\right]=0. 
\label{eq:stab}
\end{align}
This means that the 2D von Laue conditions are
satisfied if and only if the 3D ones are satisfied. 
Furthermore, the nontrivial local stability condition for the
3D~\cite{Perevalova:2016dln} and 2D ~\cite{Freese:2021czn} pressure
and shear force distributions can be considered.
The integrand in the last equation of Eq.~\eqref{eq:Abel_PS} is just
the 3D local stability condition, which implies the 2D local stability
condition on the LF~\cite{Panteleeva:2021iip} 
\begin{align}
\frac{2}{3}s(r)+p(r) >0 \iff
\frac{1}{2}\mathcal{S}(x_{\perp})+\mathcal{P}(x_{\perp}) >0. 
\label{eq:loc_stab}
\end{align}
This positivity is guaranteed by the fact that the Abel image of a
positive function is also positive and vice versa. 
This implies that the positivity of the 3D local stability condition 
is equivalent to the 2D ones. Thus, it allows one
directly to connect the 3D mechanical radius to the 2D one 
\begin{align}
\langle x^{2}_{\perp} \rangle_{\mathrm{mech}} = \frac{\int
  d^{2}x_{\perp} x^{2}_{\perp} \left(\frac{1}{2}\mathcal{S}(x_{\perp})
  +\mathcal{P}(x_{\perp})\right)}{\int d^{2}x_{\perp}
  \left(\frac{1}{2}\mathcal{S}(x_{\perp})
  +\mathcal{P}(x_{\perp})\right)} = \frac{4D(0)}{\int^{0}_{-\infty}dt
  D(t)} = \frac{2}{3}\langle r^{2} \rangle_{\mathrm{mech}}. 
\label{eq:mech}
\end{align}

To specify the meaning of the pressure and shear force distributions,
one considers the notion of the normal and tangential force fields
that are just the eigenvalues of the stress tensor contracted with the
unit radial $\bm{e}_{r}$ and tangential $ \bm{e}_{\phi}$ vectors, 
respectively. Then the 3D and the 2D force fields on the BF and LF
can be obtained by 
\begin{align}
& F_{n}(r)=4\pi
  r^{2}\left[\frac{2}{3}s(r)
  +p(r)\right], 
  \ \ \ F_{t}(r)= 4\pi
  r^{2}\left[-\frac{1}{3}s(r)
  +p(r)\right],   \cr
&F_{n}^{(2D)}(x_{\perp})=2\pi
  x_{\perp}\left[\frac{1}{2}\mathcal{S}(x_{\perp})
  +\mathcal{P}(x_{\perp})\right], 
  \ \ \ F_{t}^{(2D)}(x_{\perp})= 2\pi
  x_{\perp}\left[-\frac{1}{2}\mathcal{S}(x_{\perp})
  +\mathcal{P}(x_{\perp})\right].  
\label{eq:2Dforce}
\end{align}
The $D$-term can be also obtained in terms of the 3D and 2D pressure
and shear force distributions as  
\begin{align}
D(0) &= - \frac{4m}{15} \int d^{3}r \, r^{2} s(r)= 
  m \int d^{3}r \, r^{2} p(r)  \cr
 &= - m \int d^{2}x_{\perp} x^{2}_{\perp} \mathcal{S}(x_{\perp})= 4
  m \int d^{2}x_{\perp} x^{2}_{\perp} \mathcal{P}(x_{\perp}). 
\end{align}
This indicates that the 2D Abel images show equivalently the
mechanical structure of the nucleon as the 3D distributions do. 

\section{EMT  distributions within the chiral
  quark-soliton model}
\label{sec:3} 
We now present the results for the 2D distributions in the IMF. In
particular, we visualize the 2D EMT distributions that are derived
from the 3D ones by the Abel transform. We will use the results for
the 3D distributions in the BF, which were already obtained within the
framework of the $\chi$QSM~\cite{Goeke:2007fp, Kim:2020nug}.
Since Refs.~\cite{Goeke:2007fp, Kim:2020nug} showed in detail how the
EMT distributions can be calculated in the $\chi$QSM, we will only
mention briefly the model before we present the results. The $\chi$QSM
is known to be a pion mean-field approach to understand the structure
of the lowest-lying light and singly-heavy
baryons~\cite{Diakonov:1987ty, Christov:1995vm, Diakonov:1997sj,
  Diakonov:2010tf, Yang:2016qdz, Kim:2018cxv}. The model has been very
successful in describing various properties of baryons. For example,
the pressure distribution of the nucleon and its $D$-term form factor
obtained from the $\chi$QSM~\cite{Goeke:2007fp, Polyakov:2018zvc} were
in good agreement with recent data extracted from the 
experiments for deeply virtual Compton
scattering~\cite{Burkert:2018bqq}.  In this sense, it is of great  
interest to examine the 2D distributions on the LF based on the 
$\chi$QSM. Thus, we take the results for the 3D distributions,
i.e. $\varepsilon(r)$, $\rho_{J}(r)$, $p(r)$ and $s(r)$, which were
obtained in Ref.~\cite{Kim:2020nug} to derive the 2D EMT distributions
$\mathcal{E}(x_{\perp})$, $\rho^{(2D)}_{J}(x_{\perp})$,
$\mathcal{P}(x_{\perp})$ and $\mathcal{S}(x_{\perp})$ by the Abel
transforms. Note that in this work we employ the dynamical quark mass
as $M=350$~MeV to keep consistency with Ref.~\cite{Goeke:2007fp}
instead of $M=420$ MeV that was often used to compute baryonic
observables quantitatively.   

The 2D mass distribution given in Eq.~\eqref{eq:Abel_PS} can be
reduced in the large $N_{c}$ limit~\cite{Goeke:2007fp, Kim:2020nug} to 
\begin{align}
\mathcal{E}(x_{\perp}) =2\int^{\infty}_{x_{\perp}}
  \left(\varepsilon(r) + \frac{3}{2}p(r) \right) \frac{r
  dr}{\sqrt{r^{2}-x^{2}_{\perp}}}. 
\end{align}
A detailed discussion on the validity of neglecting the contributions
from $\partial^{2}A(t)/m$ and $\partial^{2}J(t)/m$ to the mass
distribution was held in Ref.~\cite{Polyakov:2018zvc}. 
We can see in Eq.~\eqref{eq:normal} that the contribution from the
integration of $\mathcal{E}(x_{\perp})$ over $\bm{x}_{\perp}$ is
solely due to the $\varepsilon(r)$, whereas that of the pressure 
$p(r)$ vanishes because of the 3D von Laue
condition~\eqref{eq:stab}. On the other hand, 
as presented in Eq.~\eqref{eq:radius} the 2D mass radius is expressed
not only by the 3D mass radius but also by the $D$-term, 
which is slightly different from the case of the mechanical radius
given in Eq.~\eqref{eq:mech}. As already shown in
Ref.~\cite{Panteleeva:2021iip}, the expressions of 2D radii
can be understood and generalized in terms of the Mellin moments of
the Abel images. 

\begin{figure}[htp]
\centering
\includegraphics[scale=0.27]{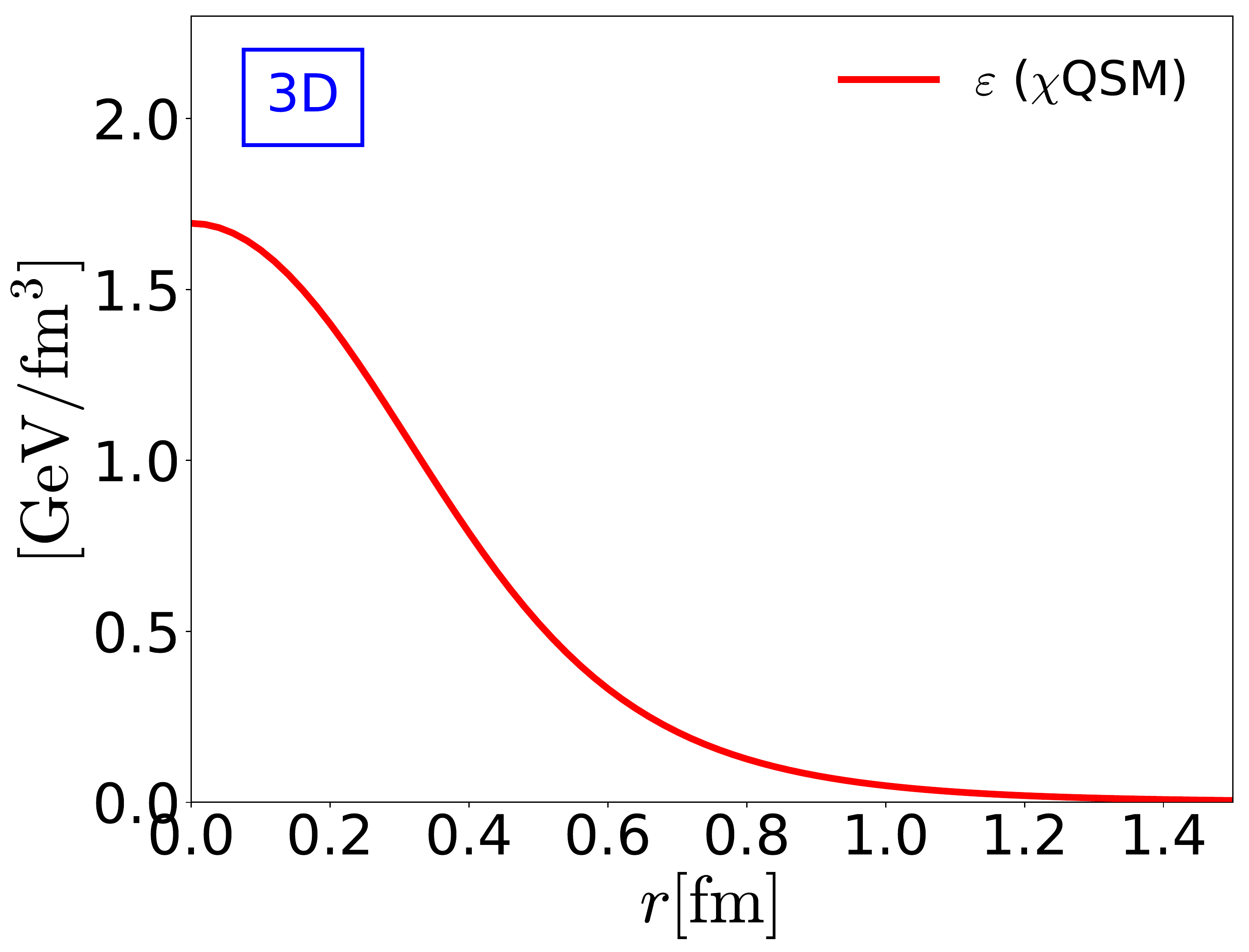}
\includegraphics[scale=0.27]{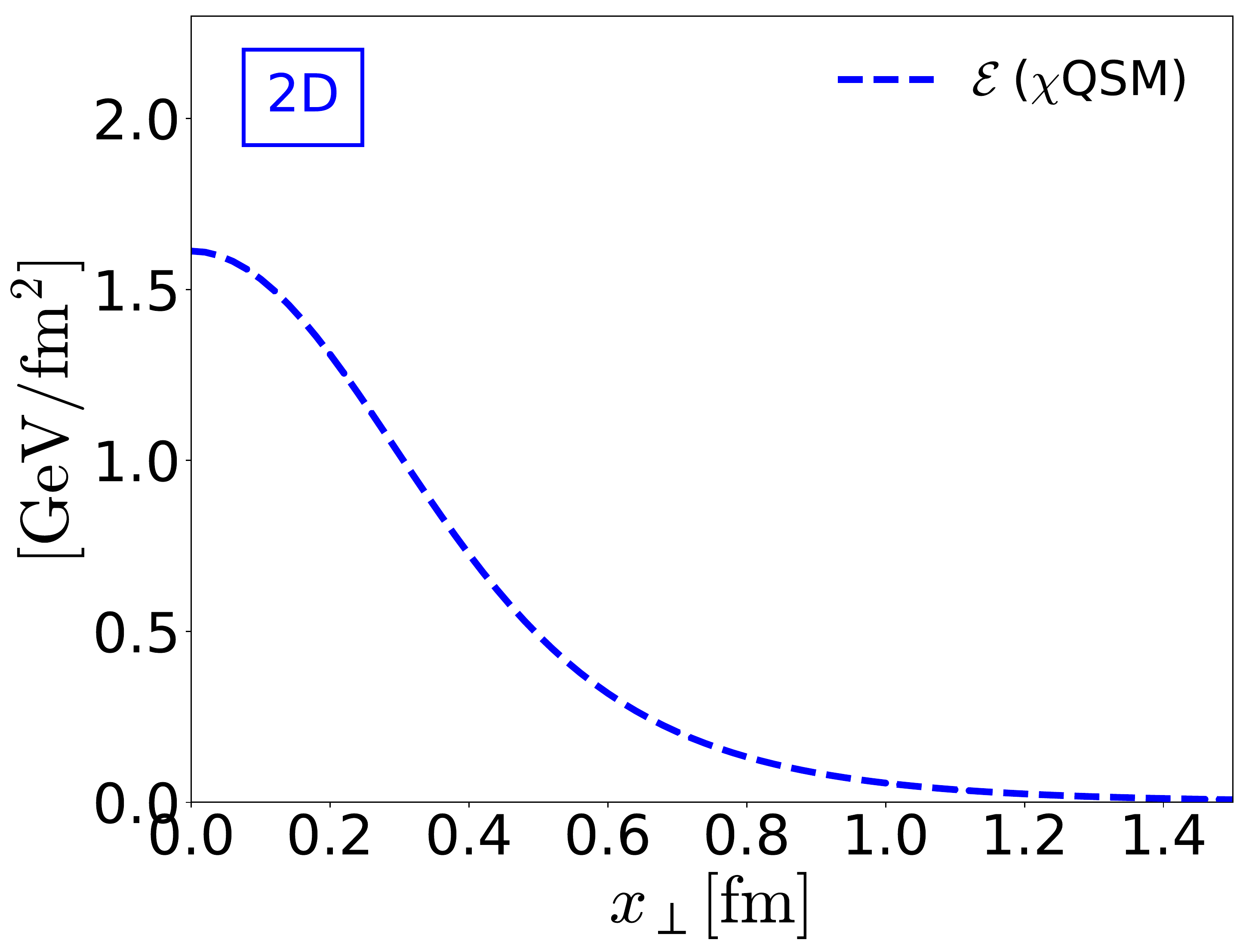}
\includegraphics[scale=0.27]{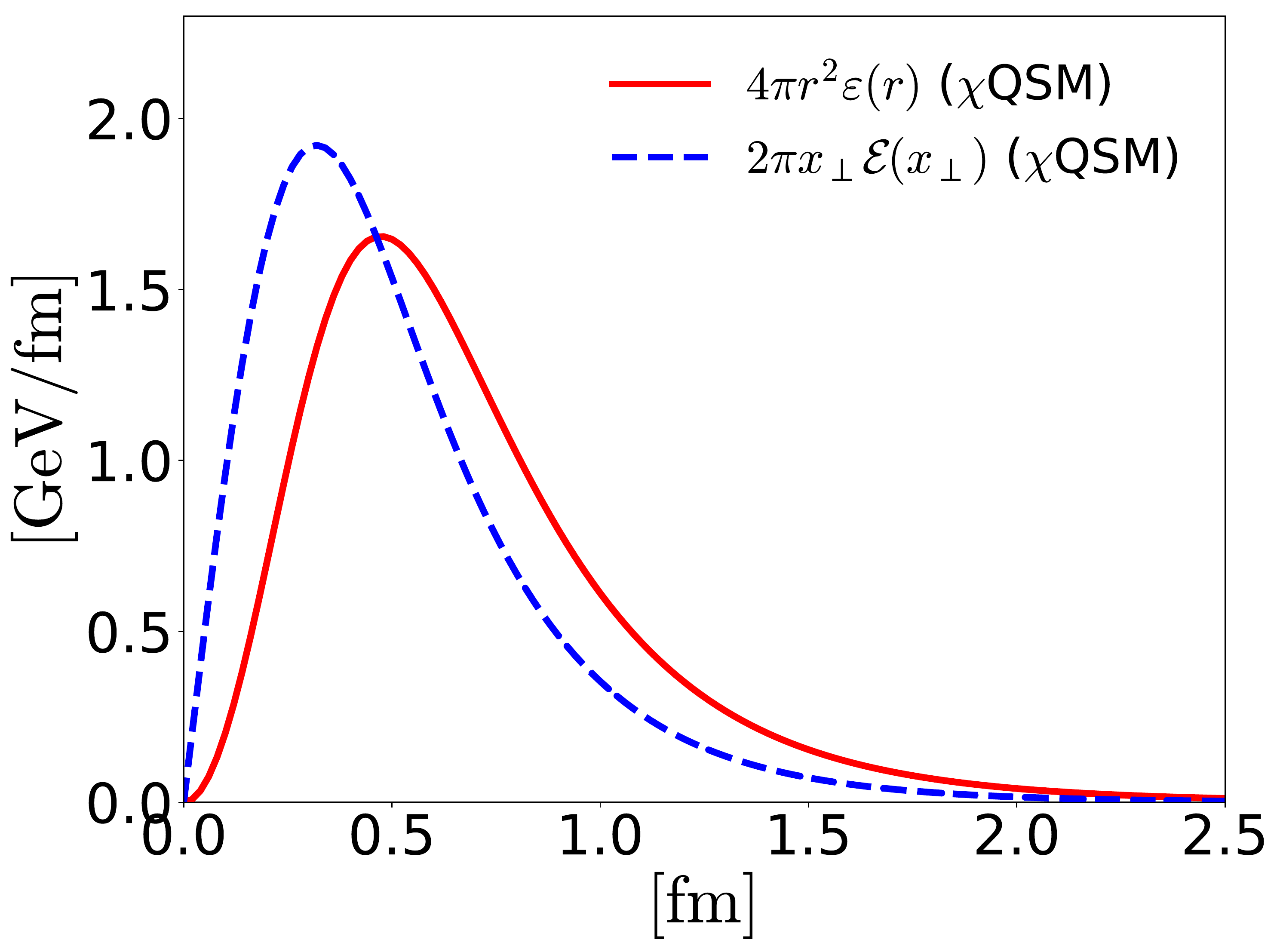}
\caption{The solid (dashed) curve in the upper left (right) panel draws
  the 3D (2D) mass distributions in the BF (IMF). The solid
  and dashed ones in the lower panel depict respectively the 3D and 2D 
  energy distributions weighted by $4\pi r^2$ and $2\pi x_\perp$.} 
\label{fig:1}
\end{figure}
We first discuss the results for the mass distributions. 
In the upper left and right panels of Fig.~\ref{fig:1}, we depict
the results for the 3D and 2D mass distributions in the BF and IMF,
respectively. The values of the mass distributions in the BF and IMF  
at the center are found to be $\varepsilon(0)=1.69\,\mathrm{GeV/fm}^{3}$
and $\mathcal{E}(0)=1.61\,\mathrm{GeV/fm}^{2}$. If we draw the 3D and
2D mass distributions weighted by $4\pi r^2$ and $2\pi x_\perp$
respectively, we can see that the 3D one exhibits a broader shape than
the 2D ones, as illustrated in the lower panel of Fig.~\ref{fig:1}.  
This indicates that the 3D mass radius should be larger than
the 2D one. The relative ratio between the 2D and 3D mean square mass
radii is given as 
\begin{align}
\frac{\langle x^{2}_{\perp} \rangle_{\mathrm{mass}}}{\langle r^{2}
  \rangle_{\mathrm{mass}}} \approx 0.6. 
\end{align}
Considering the fact that $D(0)$ should be negative,  
 $\langle x_\perp^{2} \rangle_{\mathrm{mass}}$ should be smaller than
 $\frac{2}{3}\langle r^{2} \rangle_{\mathrm{mass}}$, as expected from
 Eq.~\eqref{eq:radius}. It is of great importance to observe that
 the sign of the 3D mass distribution is preserved after the Abel
 transform. 

\begin{figure}[htp]
\centering
\includegraphics[scale=0.27]{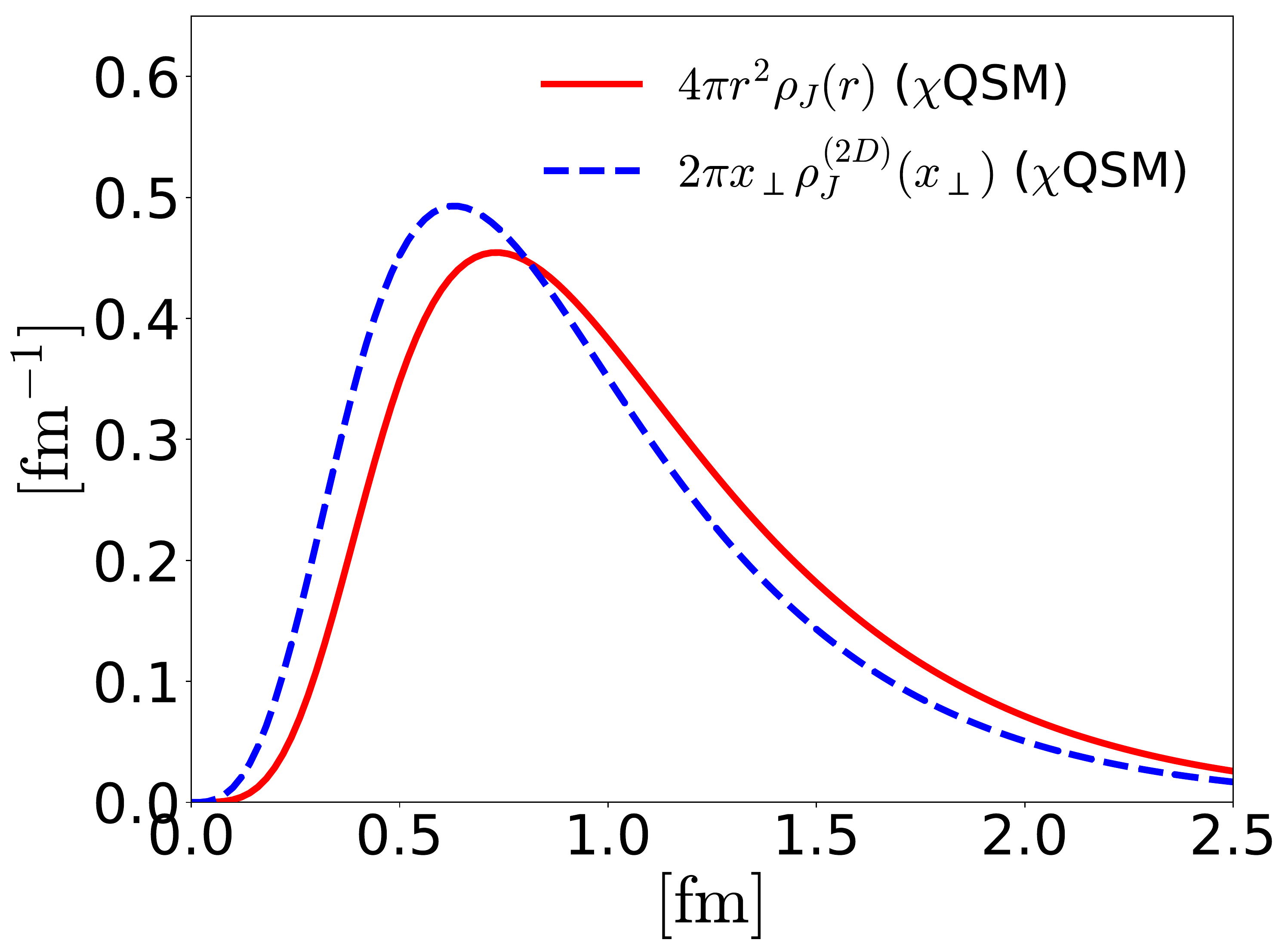}
\includegraphics[scale=0.27]{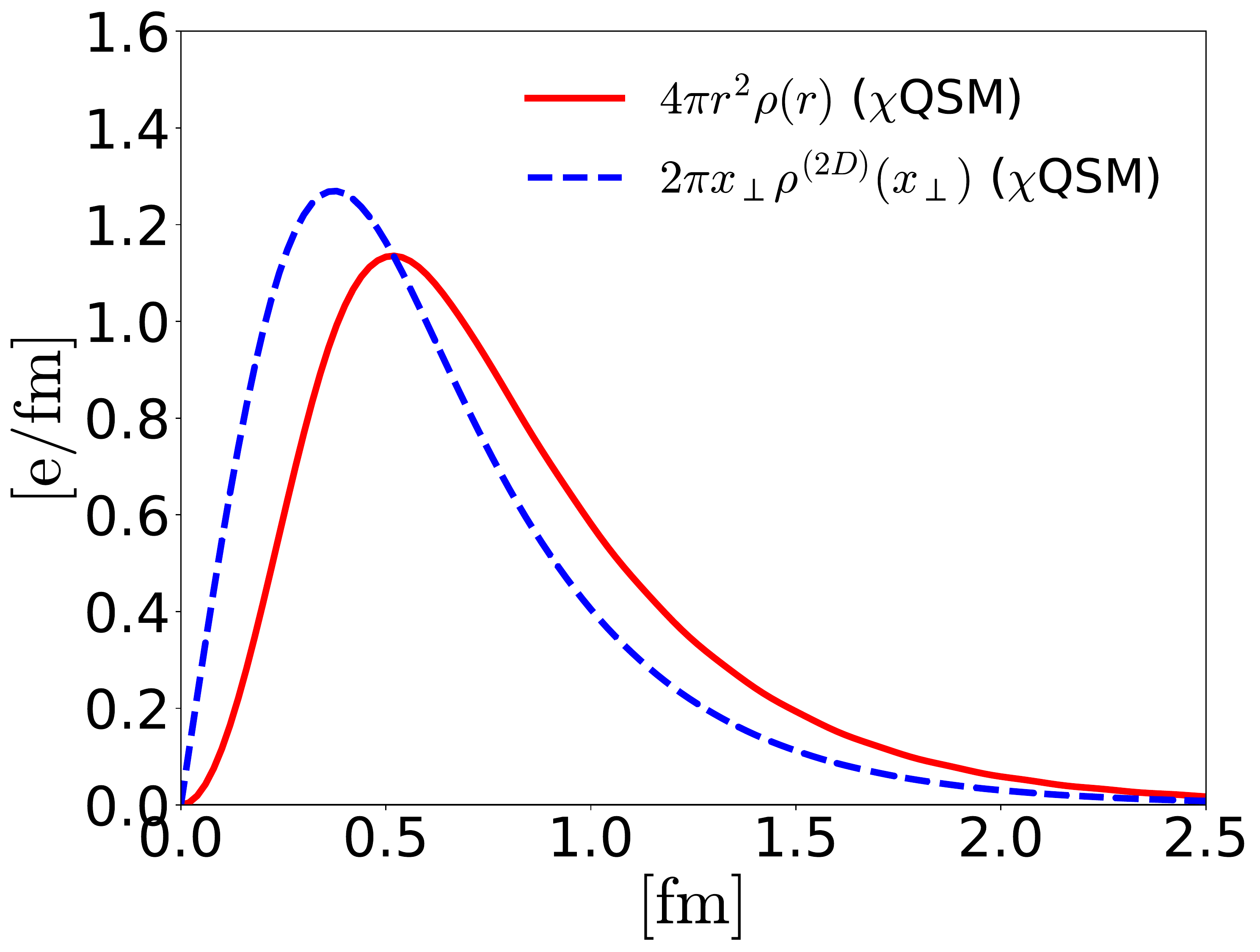}
\caption{The solid and dashed ones in the left (right) panel depict
  respectively the 3D and 2D angular-momentum (charge) distributions
  weighted by $4\pi r^2$ and $2\pi x_\perp$ in the BF and IMF,
  respectively.  
} 
\label{fig:2}
\end{figure}
The solid and dashed curves in the left panel of Fig.~\ref{fig:2} draw 
the 3D and 2D distributions for the angular momentum weighted by $4\pi
r^2$ and $2\pi x_\perp$, respectively. As shown in
Eq.~\eqref{eq:normal}, the 2D and 3D angular-momentum distributions
are normalized as 
\begin{align}
\int d^2 x_{\perp} \, \rho^{(2D)}_J(x_{\perp})=\int d^3 r \, \rho_J(r)
  = J(0) = \frac12,   
\end{align}
which is just the spin of the nucleon. Being similar to the case of 
the mass distributions, the 3D distribution is broader than the 2D
one. The 2D radius for the angular-momentum distribution is related to
the 3D one~\eqref{eq:radius}, which is smaller than the 3D radius by
the geometric factor $4/5$. Thus, the 2D distribution is not much
deviated from the 3D one as shown in the left panel of
Fig~\ref{fig:2}. For completeness, we present the 3D and 2D charge
distributions in the right panel of Fig~\ref{fig:2}. The 2D charge
radius is also related to the 3D one by the geometrical factor $2/3$,
i.e, $\langle x^{2}_{\perp}
\rangle_{\mathrm{charge}}=\frac{2}{3}\langle r^{2}  
\rangle_{\mathrm{charge}}$ in the large $N_{c}$ limit.  

\begin{figure}[htp]
\centering
\includegraphics[scale=0.27]{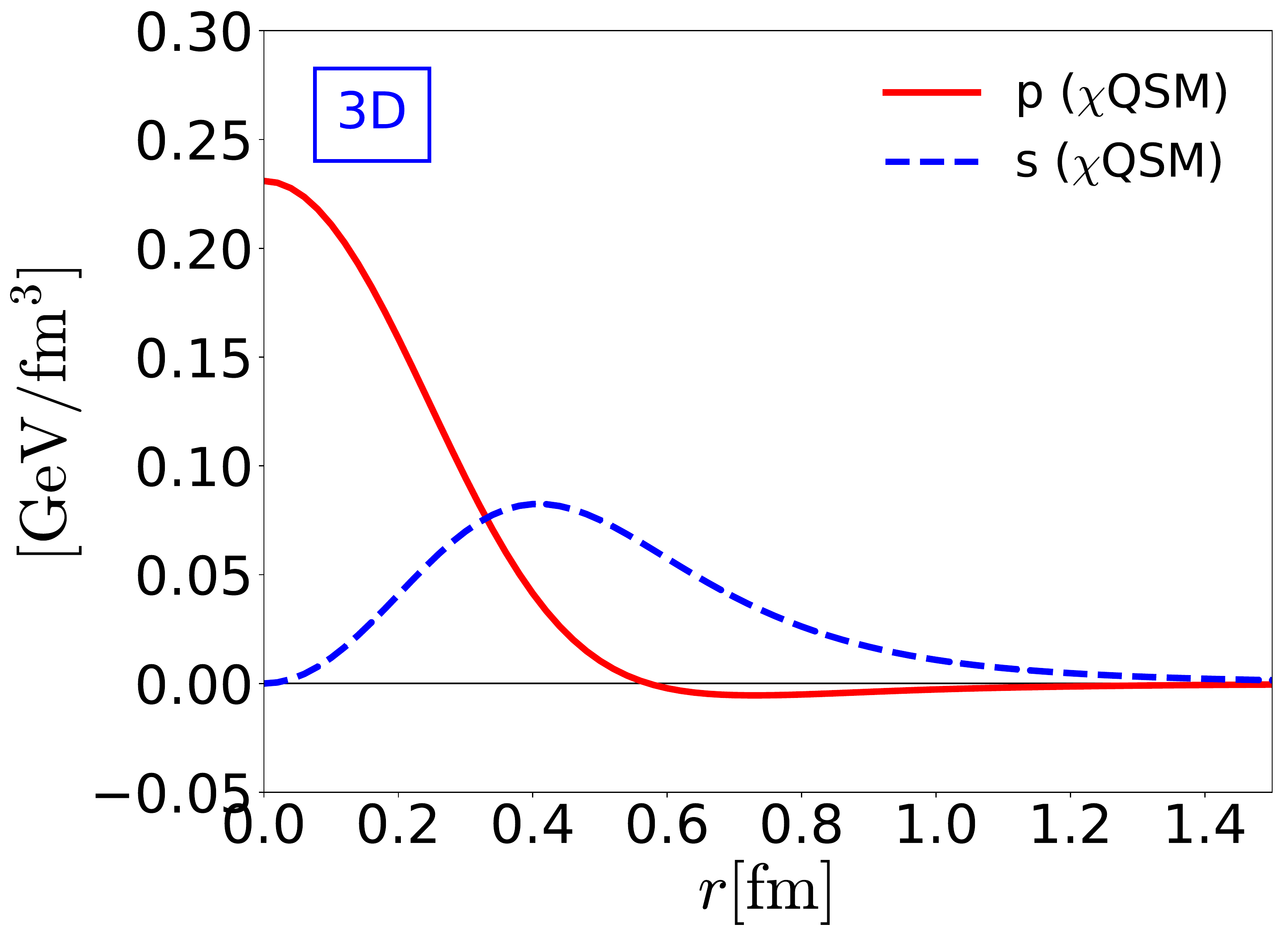}
\includegraphics[scale=0.27]{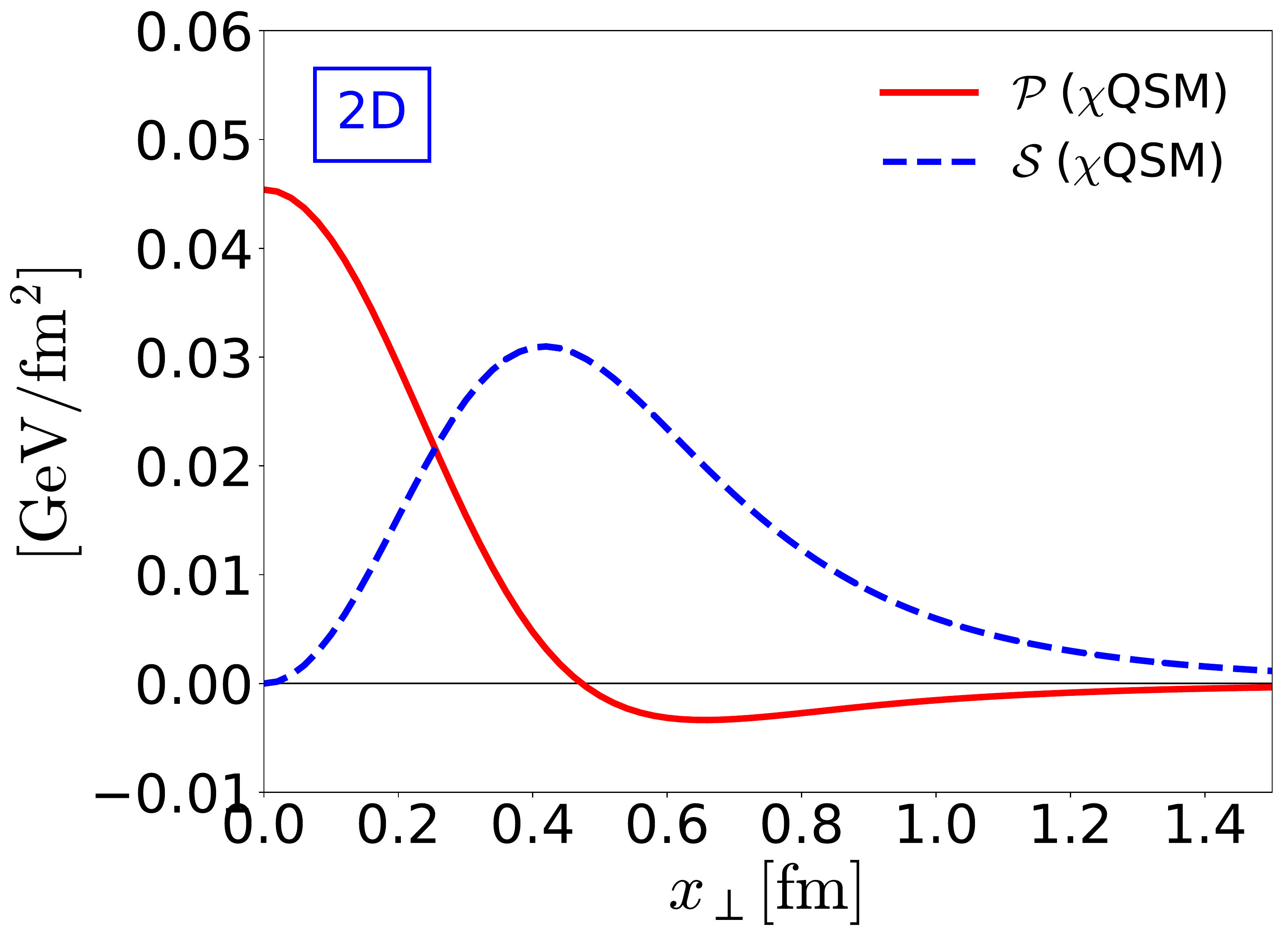}
\includegraphics[scale=0.27]{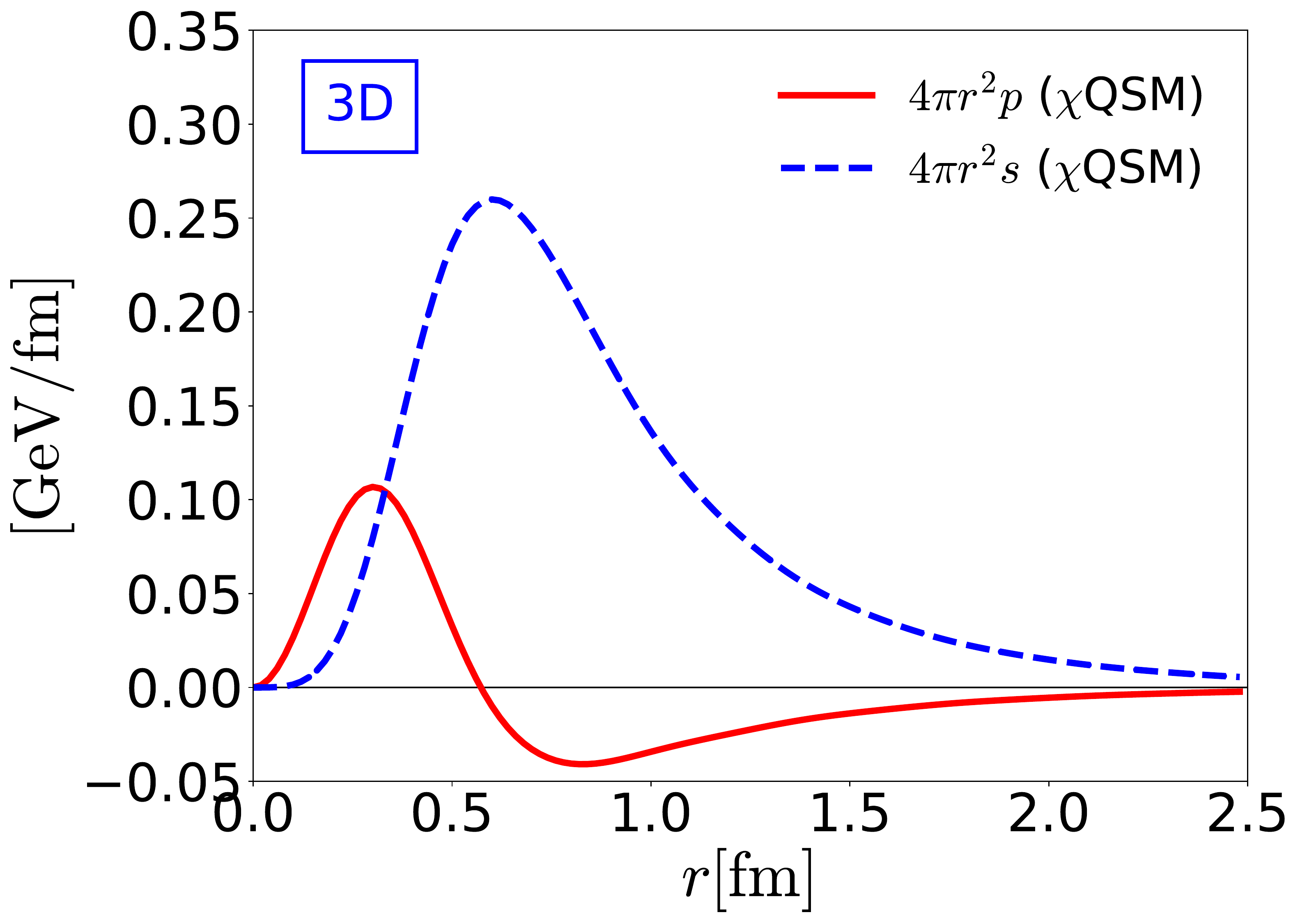}
\includegraphics[scale=0.27]{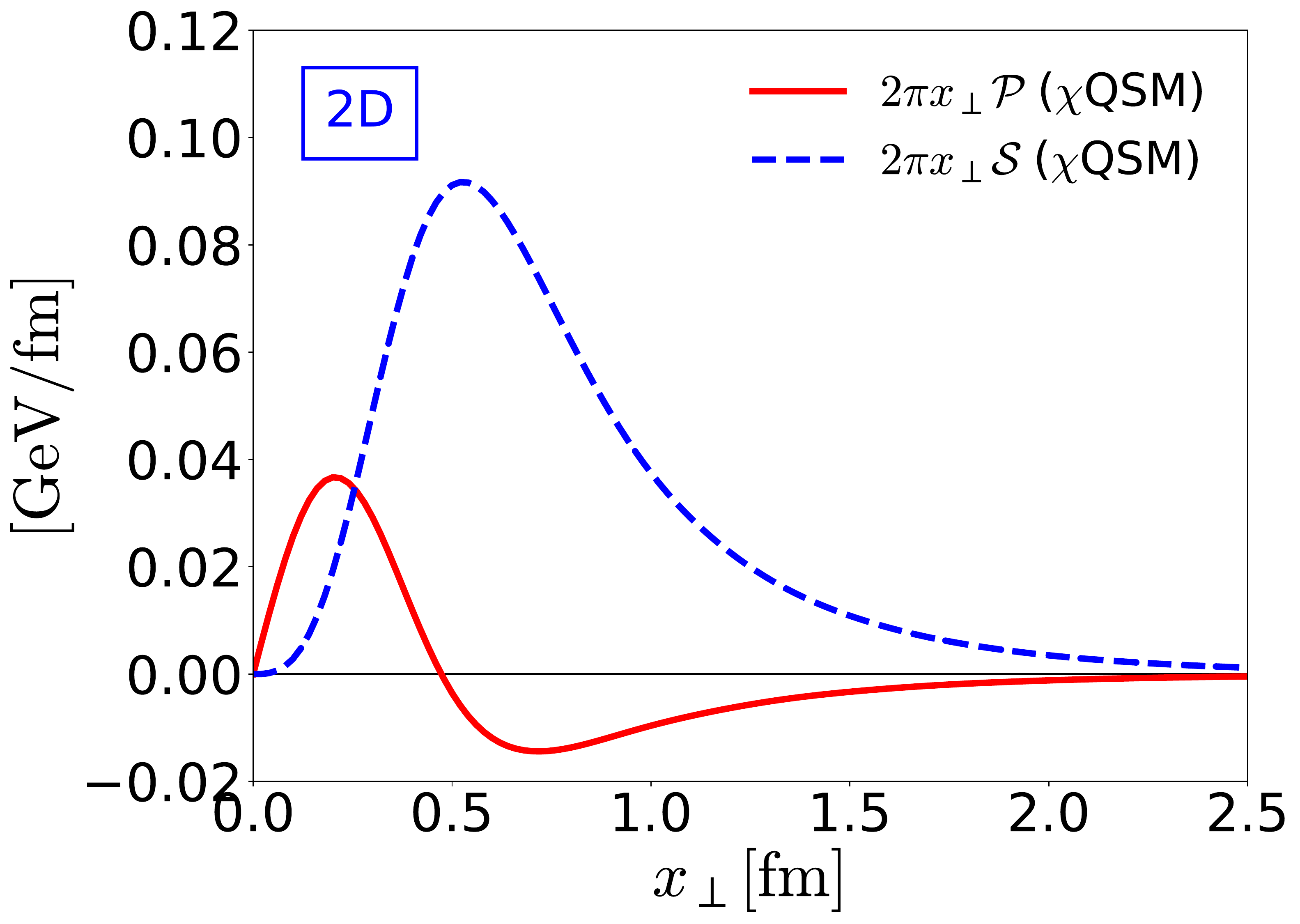}
\caption{
The  upper left (right) panel presents
  the 3D (2D) pressure and shear-force distributions in the BF
  (IMF). The solid and dashed curves denote the pressure and
  shear-force distributions respectively. 
The lower left (right) panel depicts the
3D (2D) pressure and shear-force distributions weighted by $4\pi
r^2$ $(2\pi x_\perp)$ in the BF (IMF).
}
\label{fig:3}
\end{figure}
The solid and dashed curves in the upper left (right) panel of
Fig.~\ref{fig:3} show the 3D (2D) distributions for the
pressure and shear force,  respectively. The comparison of the 2D ones
with the 3D ones provides a very important clue on the stability
condition. To comply with the 3D and 2D von Laue conditions, both the
3D and 2D pressure distributions should have at least one nodal point
$r_{0}$ and $(x_{\perp})_{0}$, respectively. The nodal points for the
3D and 2D pressure distributions are located at $r_{0}=0.57$~fm and 
$(x_{\perp})_{0}=0.47$~fm. As discussed already in
Refs.~\cite{Goeke:2007fp, Kim:2020nug}, the core part of
$p(r)$ is governed by the contribution from the level quarks and
becomes positive over the whole region of $r$, whereas the outer part
is dominated by the Dirac continuum and becomes negative. This key
feature of the pressure distribution explains how the nucleon acquires
the stability. Interestingly, the 2D pressure distribution preserves 
this significant characteristic of  the 3D one. The lower left (right)
panel of Fig.~\ref{fig:3} represents the 3D (2D) distributions for the
pressure and shear force distributions weighted by $4\pi r^{2}$ and
$2\pi x_{\perp}$, respectively, in which the key feature of the
pressure distributions is amplified.  
It is instructive to compare the ratios of the mass,
charge, angular and mechanical radii each other: 
\begin{align}
&\langle x_{\perp}^{2} \rangle_{\mathrm{mass}}  < \langle
  x_{\perp}^{2} \rangle_{\mathrm{mech}}  < \langle x_{\perp}^{2}
  \rangle_{\mathrm{charge}} < \langle x_{\perp}^{2} \rangle_{J} \;\;\;
(\mathrm{2D} \ \chi \mathrm{QSM}) \cr 
&\langle r^{2} \rangle_{\mathrm{mech}}<\langle r^{2}
  \rangle_{\mathrm{mass}}  < \langle r^{2} \rangle_{\mathrm{charge}} <
  \langle r^{2} \rangle_{J} \;\;\;\;\;\;\;\; (\mathrm{3D} \ \chi
  \mathrm{QSM}) 
\end{align}
The explicit values of the radii are listed in Table~\ref{tab:1}.
The size of the 3D mass distribution is larger than the 3D mechanical
distribution. Interestingly, this inequality of those 2D distributions
is reversed. This can be easily understood by the reason that in
the case of the 2D mass distribution in the IMF is solely determined
by the form factor $A(0)$ whereas the 3D mass distribution is related
to the form factor $A(0)$ and $D(0)$. 

\begin{figure}[htp]
\centering
\includegraphics[scale=0.27]{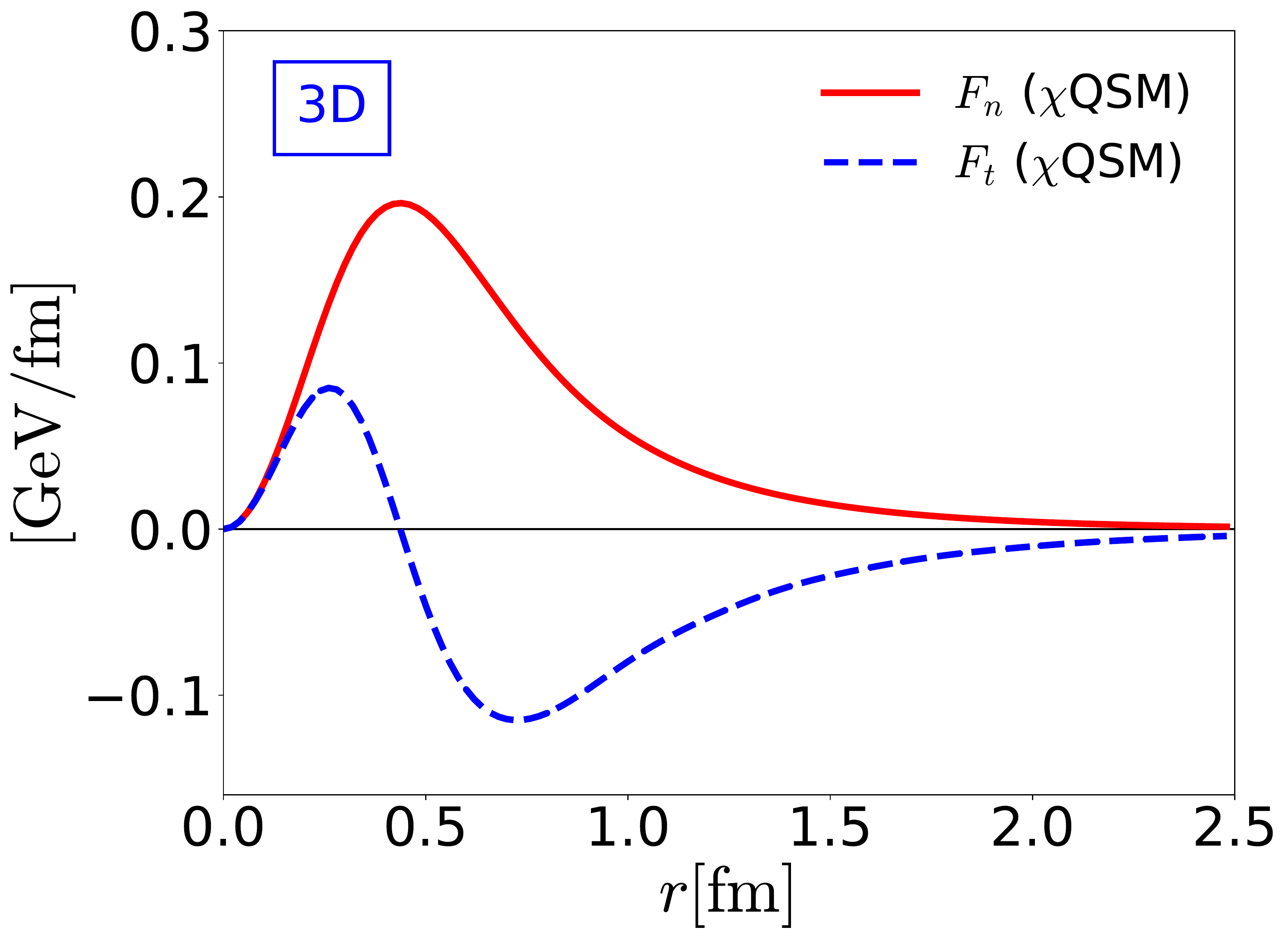}
\includegraphics[scale=0.27]{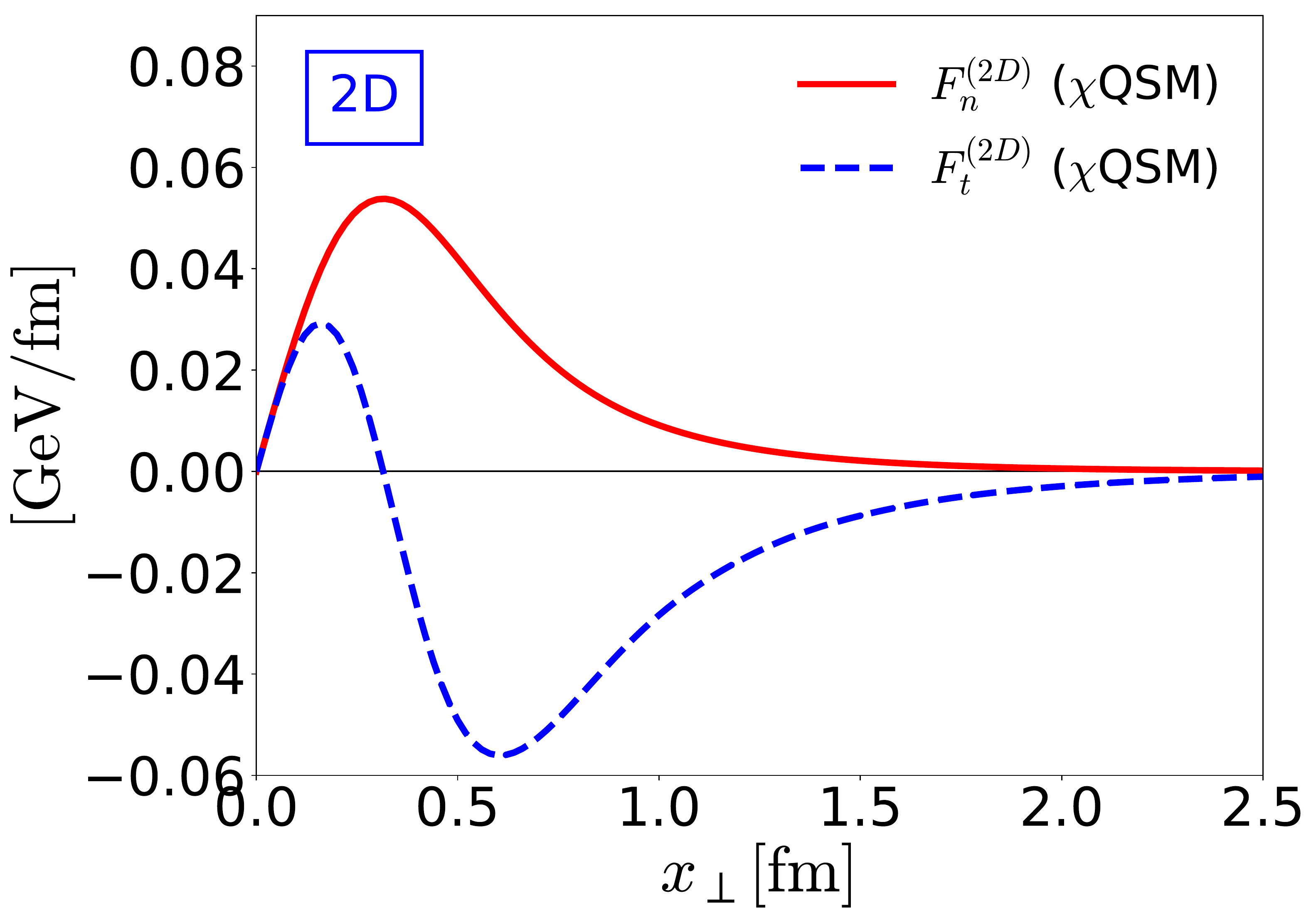}
\caption{
The solid and dashed curves in the left panel draw respectively 
  the 3D normal and tangential force fields, i.e., $F_{n}(r),\,
  F_{t}(r)$) in the BF, whereas those in the right panel 
depict respectively their counterparts $F^{(2D)}_{n}(x_{\perp})$ and 
$F^{(2D)}_{t}(x_{\perp})$ in the IMF.} 
\label{fig:4}
\end{figure}
In the left (right) panel of
Fig.~\ref{fig:4}, the solid and dashed curves depict the 3D (2D) 
distributions for the normal and tangential forces,
respectively. Indeed, the normal force field complies with the local 
stability condition~\eqref{eq:loc_stab} and the tangential force 
satisfies the von Laue condition for the 1D
subsystem~\eqref{eq:stab}. Due to this, both the 2D and 3D tangential  
forces have at least one nodal point, which means that the direction
of vector fields is reversed at this point. Note that the 3D and 2D
tangential forces change its direction at $r=0.44\, \mathrm{fm}$ and 
$x_{\perp}=0.32\, \mathrm{fm}$.
\begin{figure}[htp]
\centering
\includegraphics[scale=0.8]{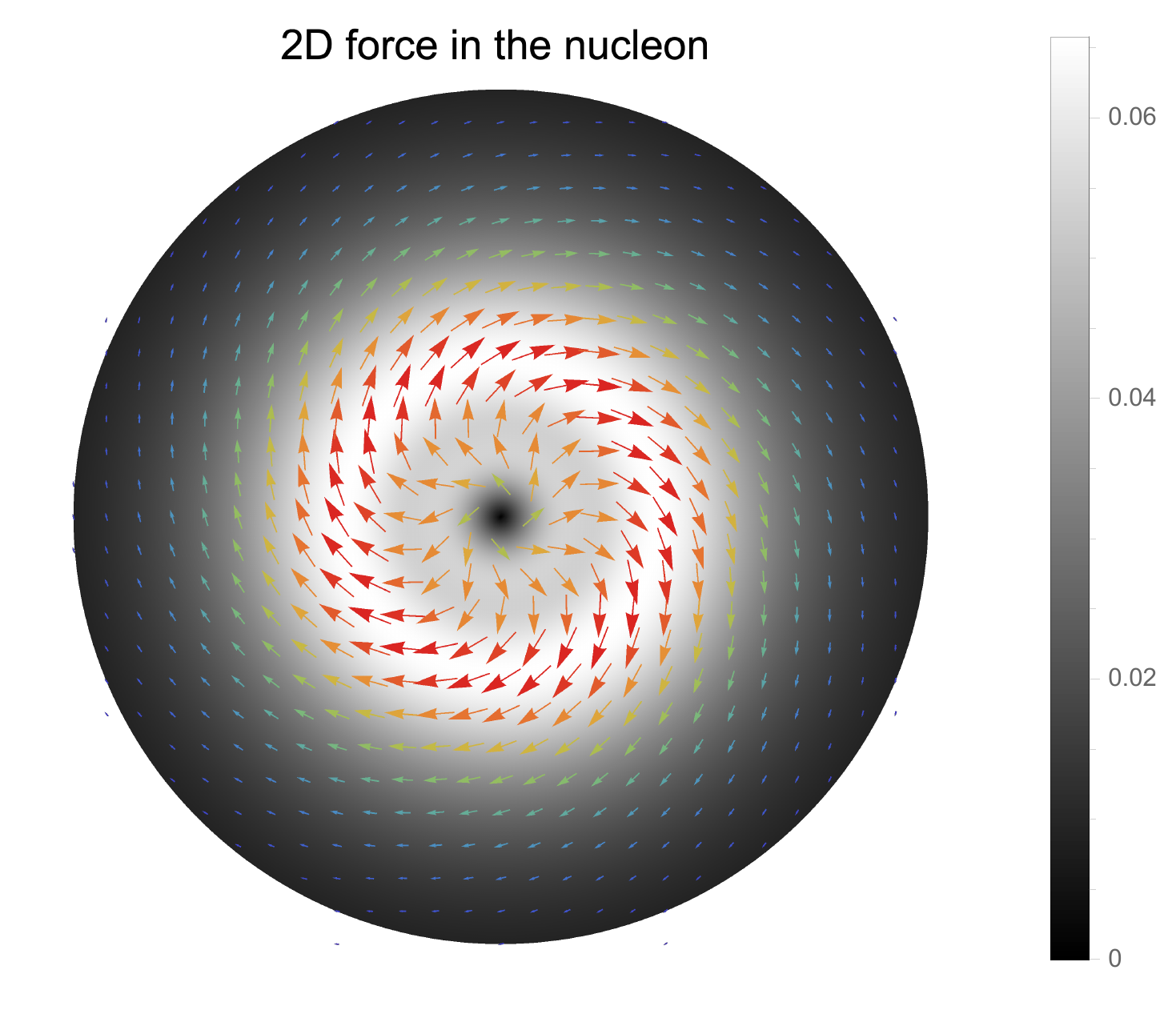}
\caption{2D visualization of the strong force field as a vector field
  inside the nucleon in the unit of $\mathrm{GeV\cdot fm^{-1}}$.
} 
\label{fig:5}
\end{figure}
 To see more clearly, the total strong force field is 
visualized in Fig.~\ref{fig:5}. The inner part of the strong forces is
governed by the normal force, whereas the outer part is dominated by
the tangential force, which ensures the nucleon to be stable. The
dominance of $F_t$ at large distances can be obtained in the
model-independent way if one uses the large $r$ behavior of the
pressure and shear force distributions derived in
Ref.~\cite{Alharazin:2020yjv}. 

\begin{table}[htp]
\setlength{\tabcolsep}{5pt}
\renewcommand{\arraystretch}{1.5}
\caption{Various observables of the EMT distributions for the nucleon
  in both the IMF and BF are listed: the energy distributions at the
  center $(\varepsilon(0), \, \mathcal{E}(0))$, the pressure
  distributions at the center $(p(0), \, \mathcal{P}(0))$, nodal
  points of the pressures $((x_{\perp})_{0},\, r_{0})$ and the mean
  square radii of the mass, angular momentum, mechanical and charge
  ($\langle r^{2} \rangle, \, \langle x^{2}_{\perp} \rangle$).} 
\begin{tabular}{ c c c | c c c c c } 
\hline
\hline{}
\centering
  $\mathcal{E}(0)$ ($\mathrm{GeV/fm}^{2}$)  & $\mathcal{P}(0)$
   ($\mathrm{GeV/fm}^{2}$)
  & $(x_{\perp})_{0}$ ($\mathrm{fm}$) & $\langle x_{\perp}^{2}
 \rangle_{\mathrm{mass}}$($\mathrm{fm}^{2}$) & $\langle
 x_{\perp}^{2} \rangle_{J}$ ($\mathrm{fm}^{2}$)
  & $\langle x_{\perp}^{2} \rangle_{\mathrm{mech}}$
    ($\mathrm{fm}^{2}$) & $\langle x_{\perp}^{2}
                          \rangle_{\mathrm{charge}}$
                          ($\mathrm{fm}^{2}$)  \\  
\hline
$1.61$ & $0.045$ & $0.47$ & $0.39$ & $1.19$ & $0.42$ & $0.58$  \\
\hline \hline
 $\varepsilon(0)$ ($\mathrm{GeV/fm}^{3}$)  & $p(0)$
 ($\mathrm{GeV/fm}^{3}$) & $r_{0}$ ($\mathrm{fm}$) & $\langle r^{2}
  \rangle_{\mathrm{mass}}$ ($\mathrm{fm}^{2}$)
  & $\langle r^{2}  \rangle_{J}$
 ($\mathrm{fm}^{2}$)
  & $\langle r^{2} \rangle_{\mathrm{mech}}$ 
($\mathrm{fm}^{2}$) &$\langle r^{2} \rangle_{\mathrm{charge}}$ 
($\mathrm{fm}^{2}$)  \\  
\hline
 $1.69$ & $0.231$ & $0.57$& $0.66$ & $1.49$ & $0.63$ & $0.86$  \\ 
 \hline
\hline
\end{tabular}
\label{tab:1}
\end{table}
In Table~\ref{tab:1}, we list the numerical results for various
observables such as the energy and pressure densities at the center of
the nucleon both in the BF and IMF. The explicit value of the nodal
points are also given. We find from the values listed in
Table~\ref{tab:1} that the magnitudes of these observables in the 2D
LF are consistently smaller than those in the 3D BF.  
\section{Summary and conclusion}
The present work aimed at investigating the equivalence between the EMT
distributions in the three-dimensional Breit frame and those on the
light front by the Abel transforms, based on the chiral quark-soliton
model. We extended the study of Ref.~\cite{Panteleeva:2021iip},
considering the energy and spin distributions both in the Breit frame
and on the light front. We also discussed the force fields inside a
nucleon on the light front. The two-dimensional global
and local stability conditions on the light front are shown to be
equivalent to the three-dimensional ones and vice versa. We compared
the results for the two-dimensional energy, spin, pressure, and
shear-force distributions with the corresponding Abel images on the
light front. The main conclusion of the present work is that while the
three-dimensional distributions in the Breit frame have only the
quasi-probabilistic meaning, it still provides intuitive understanding
of the nucleon internal structure, since the Abel transforms of such
distributions take over all essential physics from the
three-dimensional cases and preserve the global and 
local stability conditions. In this regard, it is of great importance
to investigate the Abel images for other three-dimensional
densities for the nucleon and other baryons.  

While the EMT distributions for the nucleon with spin 1/2
are projected onto the light-front transverse plane by the Abel
transform, we have to employ the more general Radon transform to
examine the two-dimensional light-front distributions for hadrons with
higher spin ($S>1/2$)~\cite{Kim:2020lrs, Polyakov:2018rew,
  Cosyn:2019aio, Polyakov:2019lbq, Panteleeva:2020ejw}. For example,
the $\Delta$ isobar, which has spin 3/2, will provide more complex
structure on the light front. It is well known that the electric
quadrupole moment of the $\Delta$ isobar measures how $\Delta$ is
deformed. This indicates that its two-dimensional transverse
distributions on the light front will shed light on how the $\Delta$
isobar is shaped. The corresponding investigation is under way.  

\begin{acknowledgments}
The authors want to express M. V. Polyakov for invaluable discussions 
and suggestions. The present work was supported by Basic Science
Research Program through the National Research Foundation of Korea
funded by the Ministry of Education, Science and Technology 
(Grant-No. 2018R1A5A1025563). J.-Y.K is supported 
by the Deutscher Akademischer Austauschdienst(DAAD) doctoral
scholarship and in part by BMBF (Grant No. 05P18PCFP1). 
\end{acknowledgments}

\end{document}